\documentclass[a4paper,10pt]{article}
\usepackage{graphicx}
\usepackage{ctable}
\usepackage{fontscale}
\usepackage{publication}
\usepackage[backend=biber,sorting=none]{biblatex}
\usepackage{algorithm}
\usepackage{algpseudocode}
\usepackage{amsmath}

\algnewcommand{\algorithmicgoto}{\textbf{go to}}%
\algnewcommand{\Goto}[1]{\algorithmicgoto~\ref{#1}}%
\title{Interrogation trajectory optimisation for Fabry-P\'erot based photoacoustic tomography}

\author[1,2,3]{Jakub Czuchnowski}
\contact{Corresponding authors: jczuchno@bu.edu, prevedel@embl.de}
\affil[1]{Cell Biology and Biophysics Unit, European Molecular Biology Laboratory, Heidelberg, Germany}
\affil[2]{Collaboration for joint PhD degree between EMBL and Heidelberg University, Faculty of Biosciences, Germany}
\affil[3]{current address: Department of Biomedical Engineering, Boston University, Boston, MA, USA}

\author[1]{Robert Prevedel}

\date{\today}

\begin{document}

\maketitle

% Abstract with justification
\begin{abstract}
    Fabry-P\'erot based photoacoustic tomography (FP-PAT) is a promising all-optical imaging modality for a wide range of preclinical and clinical applications. However, there exist several challenges in routinely applying FP-PAT in time-critical experiments. Among those, the need for spectral tuning of the laser between each scan position can severely limit the effective imaging speed. Here, we present an interrogation trajectory optimization approach which allows to increase the overall speed in a way that is independent of the type of interrogation laser used as well as the FP quality. Our approach provides a way to tackle speed degradation caused by hardware limitations and simplify the use of FP-PAT systems.
\end{abstract}

%Use this when not including an abstract to print horizontal rule and keywords.
%\noabstract

%Use either of these  to only print the keywords
%\putkeywords                                       %Prints keywords with the vertical margins and the "Keywords" identifier
%\keywordlist                                       %Provides the keywords directly

% Double-column layout
\begin{multicols}{2}

\section{Introduction}

Fabry-P\'erot based photoacoustic tomography (FP-PAT) is a promising and rapidly developing all-optical modality in the biomedical imaging field \cite{Zhang:08,Jathoul:15}. However, there are several challenges in efficiently applying FP-PAT in biological and preclinical experiments. Among those, spatial heterogeneity of the FP cavity thickness \cite{Varu:14} (that translates to spatial heterogeneity of the optimal bias wavelength) can severely limit the imaging speed by forcing spectral tuning of the laser between each scan position. While it is possible to tackle this on the hardware side either by manufacturing ultra-uniform FP interferometers \cite{Huynh:16} or by using lasers with ultra-fast wavelength scanning \cite{Saucourt:23} currently none of these are widely available to the imaging community.

Here we present an approach that bypasses these hardware limitations by optimizing the interrogation trajectory to limit the required number of tunings which allows to significantly increase the overall imaging speed. We then test this approach under realistic imaging condition and demonstrate a $\sim$ 7x increase in effective imaging speed.

\begin{figure*}[h]
\includegraphics[width=15cm]{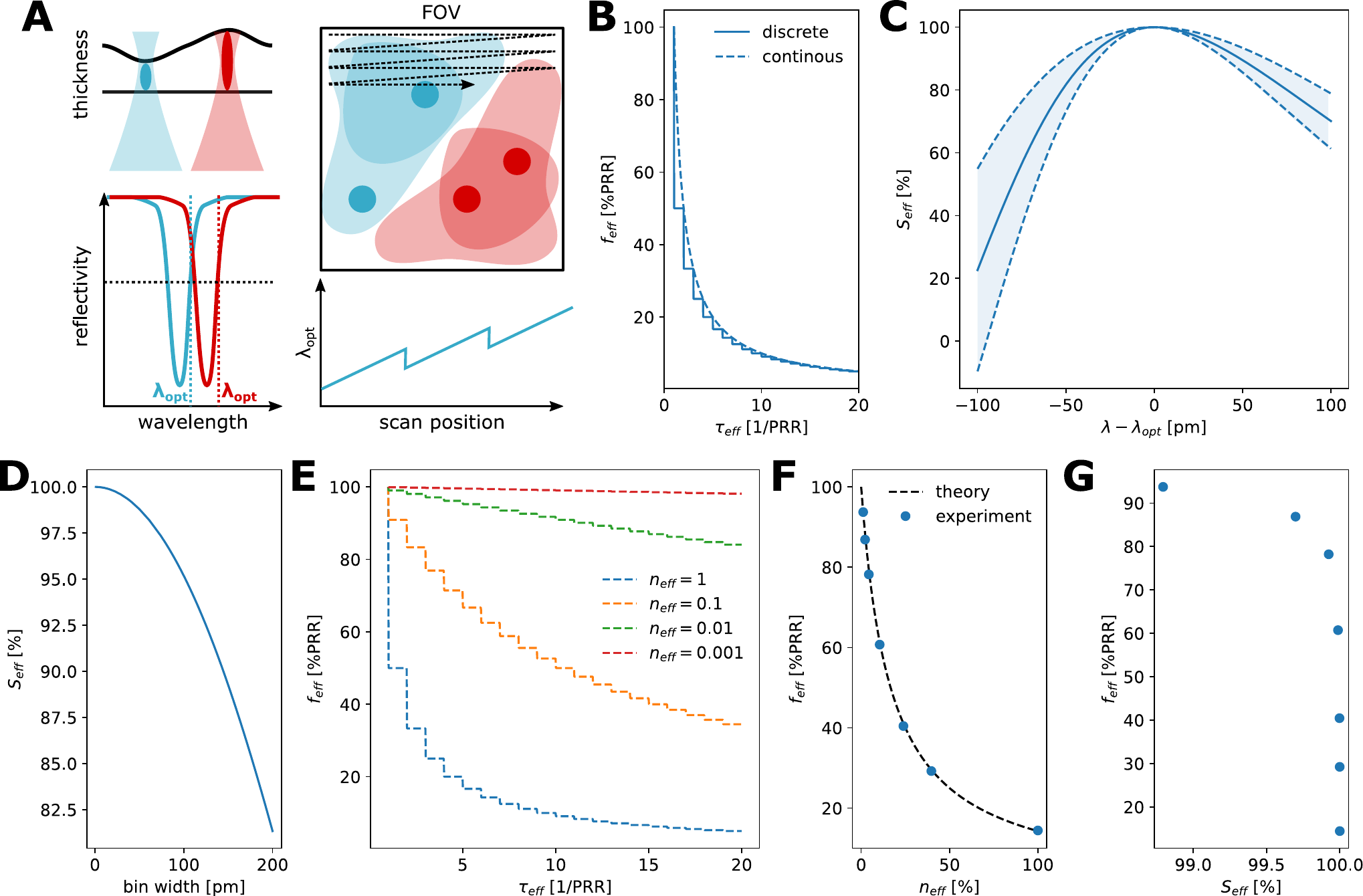}
\centering
\caption{\textbf{A} Effect of changes in FP cavity thickness on the spectral position of the transfer function and it's consequences for the optimal interrogation wavelength during raster scanning the FP surface. \textbf{B} Dependence of the effective scan frequency ($f_{eff}$) on the laser scanning lag time ($\tau_{eff}$). \textbf{C} Experimentally determined dependence of the effective sensitivity on the spectral distance of the optimal bias wavelength. \textbf{D} Effective sensitivity dependence on spectral bin width for binned interrogation. \textbf{E} Theoretical dependence on of effective scan frequency ($f_{eff}$) on the laser scanning lag time ($\tau_{eff}$) and the effective number of tunings ($n_{eff}$), see \textbf{Equation \ref{eq:f_eff}}. \textbf{F} Comparison between the experimentally  measured effective scan frequency ($f_{eff}$) and prediction from \textbf{Equation \ref{eq:f_eff}} assuming an experimentally determined $\tau_{eff}=69 \ ms$. \textbf{G} Experimentally quantified tradeoff between $f_{eff}$ and the effective sensitivity $S_{eff}$.}
\label{fig:1}
\end{figure*}

%-----------------------------------
%	SECTION 1
%-----------------------------------

\begin{figure*}[ht]
\includegraphics[width=15cm]{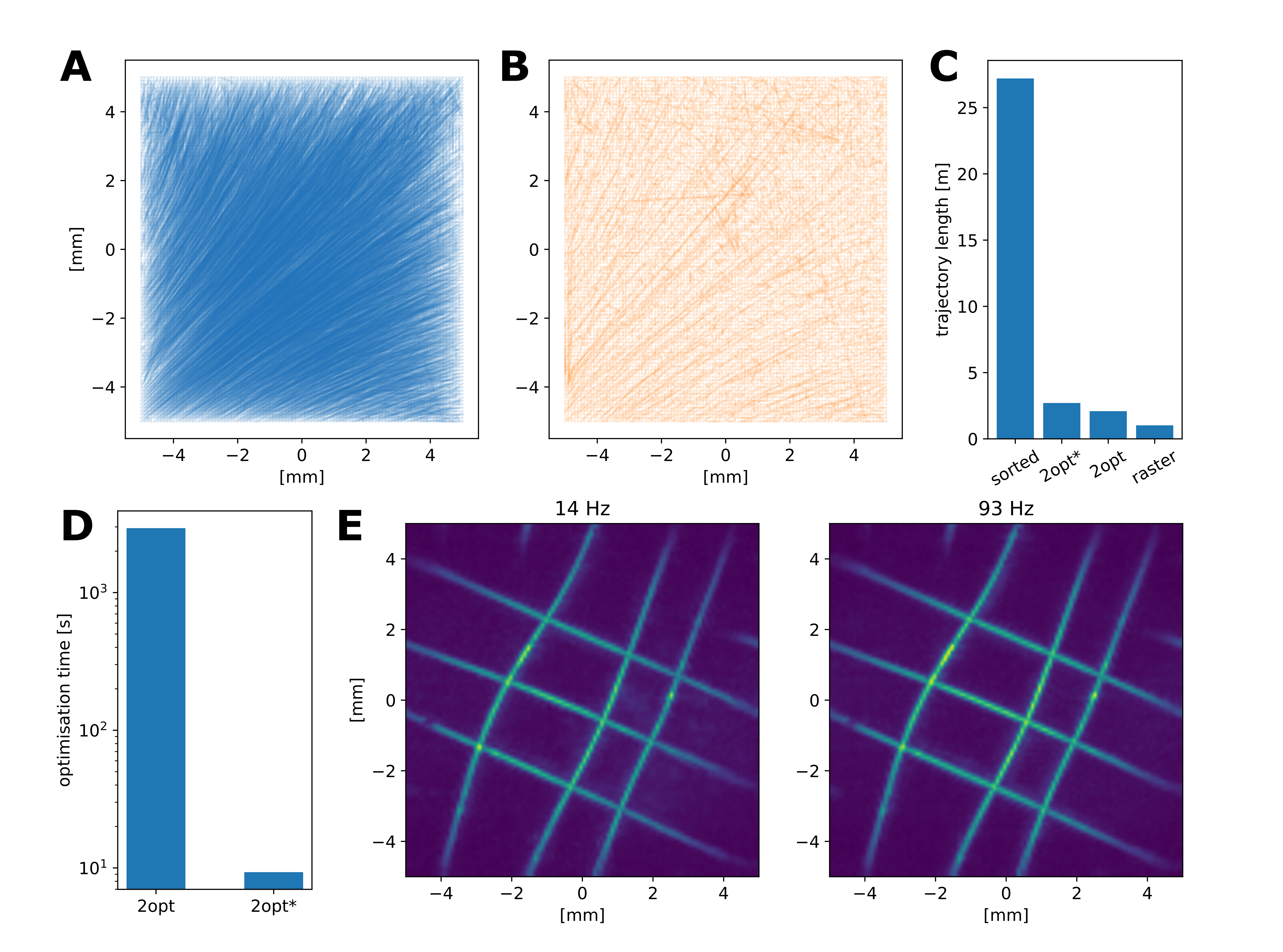}
\centering
\caption{\textbf{A} Scan trajectory across the FP surface after bias wavelength sorting. \textbf{B} Scan trajectory across the FP surface after TSS optimization. \textbf{C} Comparison of trajectory length between the sorted wavelength trajectory and the ones optimized using 2opt and the modified 2opt*, the raster scan length shows a lower bound estimation. \textbf{D} Comparison between the optimization calculation time between the 2opt and 2opt* algorithms. \textbf{E} Comparison between images taken using a normal raster scan with wavelength tuning between each step ($f_{eff}=14\ Hz$) and an image taken using an 2opt* optimized trajectory ($f_{eff}=93\ Hz$) showing no visible reduction in image quality.} 
\label{fig:2}
\end{figure*}

\section{Optimal bias wavelength binning}

Because of manufacturing imperfections, causing cavity thickness inhomogeneities, FP detectors often require tuning the bias wavelength at each scan position \cite{Zhang:08} (\textbf{Figure \ref{fig:1}A}). The effective wavelength tuning speed of the interrogation laser is key in ascertaining optimal interrogation speed. However, other parameters of the laser such as power and linewidth \cite{Varu:14} need to also be taken into consideration when choosing an interrogation laser which may require compromise on the tuning speed. Moreover, effective tuning speed is not only dependent on hardware limitations but also on the communication interface between the laser controller and the computer, as well as the control software, which may introduce significant lags. This effect can be described by a simple equation:

\begin{equation}
    f_{eff}\ [PRR]=\frac{1}{ceil(\tau_{eff})}\overset{\tau_{eff}\gg1}{\approx}\frac{1}{\tau_{eff}}
\end{equation}

where $f_{eff}$ is the effective scan frequency expressed in units of the laser pulse-repetition-rate (PRR) and $\tau_{eff}$ is the effective tuning lag of the laser expressed in 1/PRR. If $\tau_{eff} \leq 1$ the system achieves optimal PRR limited performance (\textbf{Figure \ref{fig:1}B}). However, in case of the interrogation laser used in our system the effective scan speed $f_{eff}\approx14\ Hz$ which allows to calculate the $\tau_{eff}\approx69\ ms$. This in practice allows us to achieve only a fraction of the possible speed (14\%) greatly increasing the acquisition time. This represents a major limitation and bottleneck for fast, all-optical PAT.

One possible solution to this challenge is to bin the bias wavelengths used for interrogation in order to limit the required number of tunings and thus minimize the overall effect of the tuning lag. However, binning the optimal bias wavelengths creates the problem of choosing a proper bin width. The reason being the inherent trade-off between the absolute number of discrete wavelengths needed to interrogate the sensor and the reduction of sensitivity for some points, where the used wavelength will be sub-optimal.

This loss of sensitivity is determined by the shape of the transfer function and can be characterized experimentally for each sensor (\textbf{Figure \ref{fig:1}C}). By binning these values and assuming an uniform distribution of $\lambda_{opt}$ within each bin (which is usually met in experimental conditions), the effective sensitivity trade-off with bin width can be calculated (\textbf{Figure \ref{fig:1}D}). To be able to correctly determine the required number of bins the relation needs to be established between the $f_{eff}$ and the effective number of tunings $n_{eff}$ (\textbf{Figure \ref{fig:1}E}):

\begin{equation}
    f_{eff}=\frac{1}{1+n_{eff}[ceil(\tau_{eff})-1]}
    \label{eq:f_eff}
\end{equation}

where $n_{eff}=n_t/N$, $n_t$ is the number of required tunings and $N$ is the total number of points in the scan grid. \textbf{Equation \ref{eq:f_eff}} allows to theoretically predict the behavior of our experimental system in parameter-free way requiring only the experimentally measured value of $\tau_{eff}$ (\textbf{Figure \ref{fig:1}F}). This provides a very useful tool for optimizing the system performance as the trade-off between the achieved scan speed and loss of optical sensitivity can be readily calculated without the need for experimental validation of each set of parameters.
This trade-off between $f_{eff}$ and $S_{eff}$ results in the lack of an objective optimum bin width as different applications might benefit from increases in speed and sensitivity to a varying degree. However, practically it can be observed that a minimal loss of sensitivity of around $\approx1\ \%$ can lead to a several fold increase in scan speed (\textbf{Figure \ref{fig:1}G}).

\section{Optimizing the scanning trajectory using a modified traveling salesman's problem}

While binning the bias wavelengths can dramatically reduce the number of required tunings, practically it creates a random scanning trajectory (\textbf{Figure \ref{fig:2}A}) and drastically increases the overall scanning length (\textbf{Figure \ref{fig:2}C}). In an experimental context this random access can cause galvanometric mirror instabilities and severely reduce the scan grid precision. To circumvent these we developed a framework to optimist the scan trajectory based on the 'traveling salesman problem' (\textbf{Algorithm \ref{Trajectory optimisation}}):

\begin{algorithm}[H]
  \caption{Trajectory optimization}\label{Trajectory optimisation}
  \begin{algorithmic}[1]
    \footnotesize
    \Procedure{Topt}{$\lambda_{bin}(x,y),x,y$}
    \State $full\_route=[\ ]$
    \For{$\lambda \gets min\{\lambda_{bin}\}$ \textbf{to} $max\{\lambda_{bin}\}$}
        \State $route$ = $[x[\lambda_{bin}==\lambda],y[\lambda_{bin}==\lambda]]$
        \State $optimised\_route$ = TSS($route$)
        \State $full\_route$ = concatenate($full\_route$,$route$)
        \EndFor
    \State \textbf{return} $full\_route$
  \EndProcedure
\end{algorithmic}
\end{algorithm}
where TSS stands for any 'Traveling salesman's problem solver'. As this optimization needs to be done in real-time we choose a time-efficient algorithm called '2opt' \cite{Croes:58,Flood:56} (\textbf{Algorithm \ref{2opt}}), which can efficiently reduce the length of the scan trajectory (\textbf{Figure \ref{fig:2}B,C}).

\end{multicols}
\begin{algorithm}[H]
  \caption{2opt}\label{2opt}
  \begin{algorithmic}[1]
    \footnotesize
    \Procedure{2opt}{$route$}
    \While{$improvement \ is \ made$}
      \State $best \_ distance$ =calculateTotalDistance($existing \_ route$) \label{marker}
      \For{$i\gets0$ \textbf{to} number of nodes eligible to be swapped - 1}
      \For{$k\gets i+1$ \textbf{to} number of nodes eligible to be swapped - 1}
        \State $new\_route$ = 2optSwap($existing\_route, i, k$) \Comment{\textbf{see Algorithm \ref{2optSwap}}}
        \State $new\_distance$ = calculateTotalDistance($new\_route$)
        \If{$distance_{new}<distance_{best}$}
        \State$existing \_ route= new \_ route$
        \State$best \_ distance=new \_ distance$
        \State \Goto{marker}
        \EndIf
        \EndFor
        \EndFor
    \EndWhile
    \State \textbf{return} $existing \_ route$
  \EndProcedure
\end{algorithmic}
\end{algorithm}

\begin{algorithm}[H]
  \caption{2optSwap}\label{2optSwap}
  \begin{algorithmic}[1]
    \footnotesize
    \Procedure{2optSwap}{$route, i, k$}
    \State initialize $new \_ route$ 
    \State take $route[0]$ to $route[i-1]$ and add them in order to $new \_ route$
    \State take $route[i]$ to $route[k]$ and add them in reverse order to $new \_ route$
    \State take $route[k+1]$ to end and add them in order to $new \_ route$
    \State \textbf{return} $new \_ route$
  \EndProcedure
\end{algorithmic}
\end{algorithm}
\begin{multicols}{2}

However, due to the large overall number of points in our trajectory ($\sim 10.000$) the full 2opt algorithm takes a significant amount of time to converge (\textbf{Figure \ref{fig:2}D}). To circumvent this we developed a modified 2opt algorithm (\textbf{Algorithm \ref{2opt*}}), which, while it does not ascertain an 2opt-optimal final trajectory, is efficient in rapidly reducing the overall trajectory length (\textbf{Figure \ref{fig:2}C}).

We found that even a single iteration is sufficient to achieve a stable trajectory that does not disrupt the performance of the galvo-mirrors. We validated this approach by imaging wire phantoms (see Ref \cite{Czuchnowski:21} for details) and observed no significant image quality loss (\textbf{Figure \ref{fig:2}E}).

\end{multicols}
\begin{algorithm}[H]
  \caption{2opt*}\label{2opt*}
  \begin{algorithmic}[1]
  \footnotesize
    \Procedure{2opt*}{$route,iteration \_ limit$}
    \For{$iterations\gets1$ \textbf{to} $iteration \_ limit$}
      \State $best \_ distance$ =calculateTotalDistance($existing \_ route$)
      \For{$i\gets0$ \textbf{to} number of nodes eligible to be swapped - 1}
      \For{$k\gets i+1$ \textbf{to} number of nodes eligible to be swapped - 1}
        \State $new\_route$ = 2optSwap($existing\_route, i, k$) \Comment{\textbf{see Algorithm \ref{2optSwap}}}
        \State $new\_distance$ = calculateTotalDistance($new\_route$)
        \If{$distance_{new}<distance_{best}$}
        \State$existing \_ route= new \_ route$
        \State$best \_ distance=new \_ distance$
        \EndIf
        \EndFor
        \EndFor
    \EndFor
    \State \textbf{return} $existing \_ route$
  \EndProcedure
\end{algorithmic}
\end{algorithm}
\begin{multicols}{2}
%De brabandare 2017

\section{Discussion}

Fabry-Pérot pressure sensors are promising and rapidly developing type of all-optical detectors for photoacoustic imaging. However, they require complex interrogation schemes with the use of tuneable lasers which can often negatively impact their performance. In this work we presented a simple scheme that can allow almost pulse-repetition-rate limited performance even for imperfect FP interferometers and considerable lag times in the laser tuning. With the expected further improvements in the PRR of the excitation lasers, it will thus be increasingly difficult to match the required short tuning time for the interrogation lasers. Our work provides an innovative approach and framework to overcome the imposed limitations and open the door for faster and more time-efficient FP-PAT implementations.

\section*{Funding}
This work was supported by the European Molecular Biology Laboratory (EMBL), the Chan Zuckerberg Initiative (Deep Tissue Imaging grant no. 2020-225346), as well as the Deutsche
Forschungsgemeinschaft (DFG, project no. 425902099).

\section*{Acknowledgments}

We acknowledge Florian Mathies, Johannes Zimmermann and Gerardo Hernandez-Sosa from InnovationLab Heidelberg as well as Karl-Phillip Strunk and Jana Zaumseil from Centre for Advanced Materials, Heidelberg University for help with manufacturing of the Fabry-P\'erot interferometers with elastic cavities used in this work. We acknowledge the Mechanical Workshop at EMBL Heidelberg for manufacturing custom opto-mechanical components required for the experimental setup.

\section*{Disclosures}

The authors declare that there are no conflicts of interest related to this article.

\end{multicols}

%\nocite{*} %This ensures all citations are shown, even when unused
\printbibliography

\newpage

%\begin{appendices}
%%    \section{Information}
 %   You may put additional information here
%\end{appendices}

\end{document}